# SYSTEM ANALYSIS AND DESIGN FOR MULTIMEDIA RETRIEVAL SYSTEMS

Avinash N Bhute[1] and B B Meshram[2]

[1,2]Department of Computer Engineering, VJTI, Matunga, Mumbai-19

## ABSTRACT

*Due to the extensive use of information technology and the recent developments in multimedia systems, the amount of multimedia data available to users has increased exponentially. Video is an example of multimedia data as it contains several kinds of data such as text, image, meta-data, visual and audio. Content based video retrieval is an approach for facilitating the searching and browsing of large multimedia collections over WWW. In order to create an effective video retrieval system, visual perception must be taken into account. We conjectured that a technique which employs multiple features for indexing and retrieval would be more effective in the discrimination and search tasks of videos. In order to validate this, content based indexing and retrieval systems were implemented using color histogram, Texture feature (GLCM), edge density and motion..*

## KEYWORDS

*Video, retrieval, audio, image, color feature, texture feature, edge, indexing,*

## 1. INTRODUCTION

Assisting a human operator to receive a required video sequence within a potentially large database is called as Video retrieval system. Visual media requires large amounts of storage and processing, so there is a need to efficiently index, store, and retrieve the visual information from multimedia database. The goal of the image indexing is to develop techniques that provide the ability to store and retrieve images based on their contents [1, 2]. This research of video retrieval began by retrieving a text i.e. a attached tag or text. The problem with this approach was manual annotation and lot of storage. Hence a novel technique of retrieving a video by the visual content of image/video is the hot area of research. Many such systems are proposed in the literature. The schematic block diagram of our model is shown in figure 1. A video retrieval system basically contains three steps, Video segmentation, Feature extraction, and video retrieval using similarity matching algorithm [4, 5].

Methods for shot segmentation or shot boundary detection usually first extract visual features from each frame, then measure similarities between frames using the extracted features, and, finally detect shot boundaries that are dissimilar [2]. In the second step the visual contents of the video clips i.e. color, shape, texture and motion is extracted depending on the application in hand. These features are then stored in a vector so that they can further be use for testing the similarity between the query clip and video clips in the database. In the last step the computed vectors in the earlier step is compared with the query clip vector feature, which is called as similarity matching algorithm. Most of these algorithms are based on distance measure. Use of these measures may guaranty the speed but the video clips retrieved may be redundant video clips. For more accuracy we proposed LSI based Similarity matching. This approach need more computation and storage but this problem can be easily overcome because of the availability cheep storage, fast and advanced processors or machines.

## 2. PROPOSED SYSTEMS

### 2.1 Systems Architecture

As in internet era most difficult task is to retrieve the relevant information in response to a query. To help a user in this context various search system/engine are there in market with different features. In web search era 1.0 the main focus was on text retrieval using link analysis. It was totally read only era. There was no interaction in between the user and the search engine i.e. after obtaining search result user have no option to provide feedback regarding whether the result is relevant or not. In web search era 2.0 the focus was on retrieval of data based on relevance ranking as well as on social networking to read, write, edit and publish the result. Due to Proliferation of technology the current search era based on contextual search. Where rather than ranking of a page focus is on content based similarity to provide accurate result to user.

The CBVR (Content Based Video Retrieval) have received intensive attention in the literature of video information retrieval since this area was started couple of years ago, and consequently a broad range of techniques has been proposed. The algorithms used in these systems are commonly divided into four tasks:

- Segmentation
- Extraction
- Selection, and
- Classification

The segmentation task splits the video into number of chunks or shots. The extraction task transforms the content of video into various content features. Feature extraction is the process of generating features to be used in the selection and classification tasks. A feature is a characteristic that can capture a certain visual property of an image either globally for the whole image, or locally for objects or regions. Feature selection reduces the number of features provided to the classification task. Those features which are assisting in discrimination are selected and which are not selected is discarded. The selected features are used in the classification task [2]. The figure 1 shows the content based video search systems with four primitive tasks.

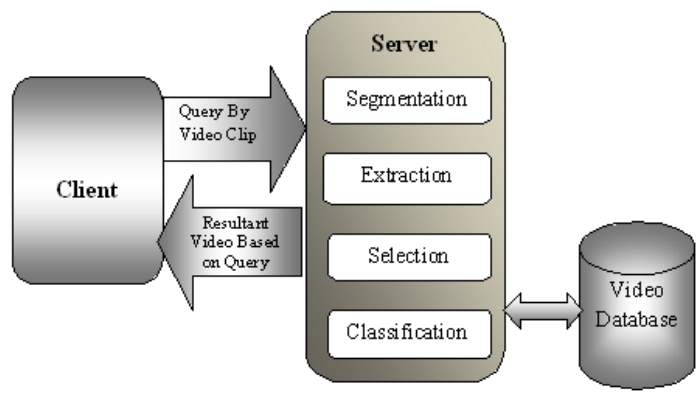

Fig.1 Content Based Video Search Systems with Primitive Tasks.

Among these four activities feature extraction is critical because it directly influence the classification task. The set of features are the result of feature extraction. In past few years, the number of CBVR systems using different segmentation and extraction techniques, which proves reliable professional applications in Industry automation, social security, crime prevention, biometric security, CCTV surveillance [1], etc.

## 2.2. Proposed Systems

Due to rapidity of digital information (Audio, Video) it become essential to develop a tool for efficient search of these media. With help of this paper we are proposing a Video Search system which will provide accurate and efficient result to a user query. The proposed system is a web based application as shown in fig.1 which consists of following processing:

1)  *Client Side Processing*

>   From client machine user can access the Graphical User Interface of the system. User can access and able to perform three tasks:

>   a)  Register the video
>   b)  Search the video
>   c)  Accept the efficient result from server.

2)  *Server Side Processing*

>   The core processing will be carried out at server side to minimize the overhead on client. Client will make a request for similar type of videos by providing query by video clip. On reception of this query by video clip, server will perform some processing on query video as well as on videos in its database and extract the video which are similar to query video. After retrieving the similar videos from the database, server will provide the list to the client in prioritized order. To accomplish this following operations are carried out at the server.

>   - Video segmentation and keyframe extraction
>   - Feature Extraction
>   - Matching keyframe features with feature database.
>   - Index the keyframes.
>   - Provide the resultant video to client in prioritized order.

## 3. SYSTEMS ANALYSIS

In this section we present system analysis of proposed systems. System analysis model defines user requirements, establishes basis of system design and defines set of validation requirements needed for testing implemented system. Beginning once system requirements have been analyzed and specified, System design is the first of three technical activities design, code generation, and test—that are required to build and verify the system. Each activity transforms information in a manner that ultimately results in validated proposed System.

At the core of the model lies the data dictionary—a repository that contains descriptions of all data objects of system. Three different diagrams surround the core. The entity relation diagram (ERD) depicts relationships between data objects. The ERD is the notation that is used to conduct the data modeling activity. The attributes of each data object noted in the ERD can be described using a data object description.

The data flow diagram (DFD) serves two purposes- to provide an indication of how data are transformed as they move through the system and to depict the functions and sub functions that transform the data flow. The DFD provides additional information that is used during the analysis of the information domain and serves as a basis for the modeling of function. A description of each function presented in the DFD is contained in a process specification (PSPEC). The state transition diagram (STD) indicates how the system behaves as a consequence of external events. To accomplish this, the STD represents the various modes of behavior (called states) of the system and the manner in which transitions are made from state to state. The STD serves as the basis for behavioral modeling

## 3.1 Use Case Diagram

A use-case is a scenario that describes how System is to be used in a given situation. To create a use-case, the analyst must identify the different types of people (or devices) that use the system or product. These actors actually represent roles that people (or devices) play as the system operates. An actor is anything that communicates with the system or product and that is external to the system itself. A use case corresponds to a sequence of transaction, in which each transaction is invoked from outside the system (actors) and engages internal objects to interact with one another and with the system surroundings. Use case represents specific flows of events in the system; it is also a graph of actors, a set of use cases enclosed by a systems boundaries communication between actors and the use cases. It is shown in following figure 2.

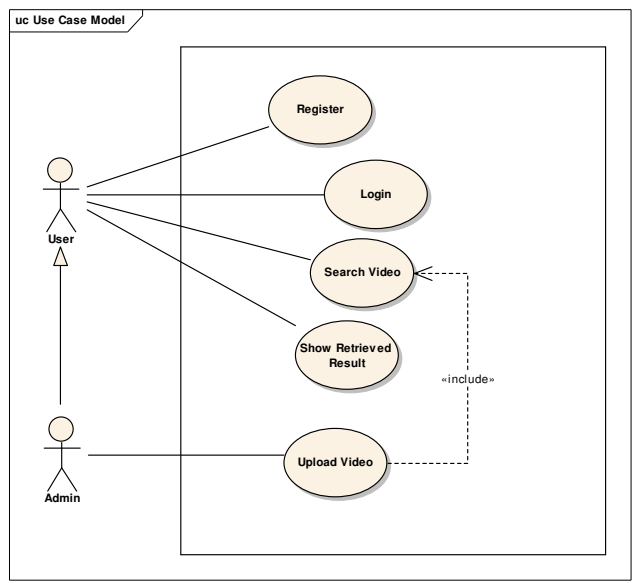

Figure 2. Use case Diagram for Video search systems

## 3.2 Class Diagram

In class diagram different classes are connected with other by means of different relations like aggregation, association, generalization etc.

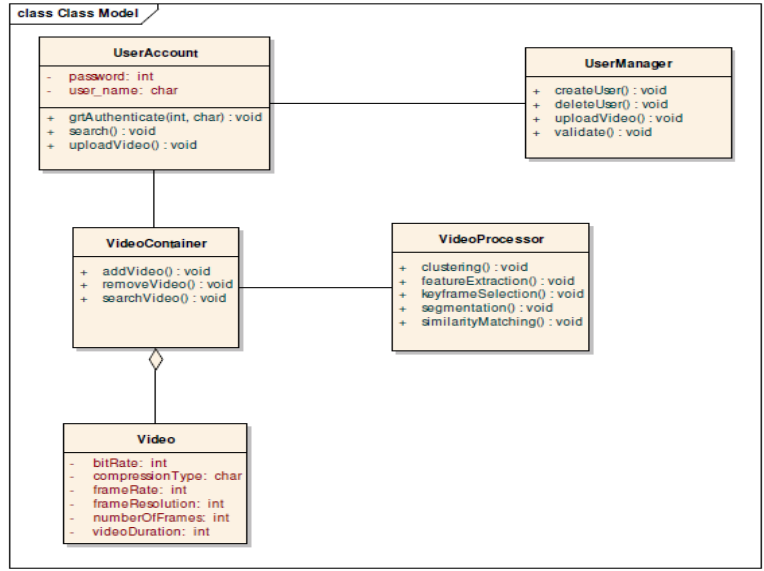

Figure3. Class Diagram for video search Systems

### 3.3 Activity Diagram

Activity diagram is basically a flow chart to represent the flow form one activity to another activity. The activity can be described as an operation of the system. So the control flow is drawn from one operation to another. This flow can be sequential, branched or concurrent. Activity diagrams deals with all type of flow control by using different elements like fork, join etc. The figure 4 is activity diagram for administrator, that means the above shown various activities are performed by admin with one initial and one final state and figure 5 is activity diagram for user, various activities are performed by user with one initial and one final state.

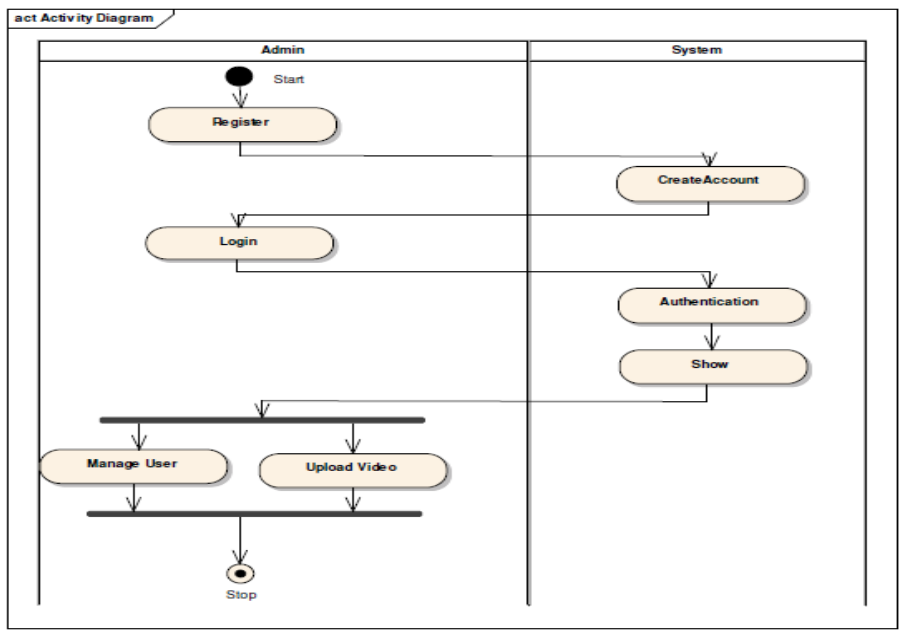

Figure 4. Activity diagram for Admin

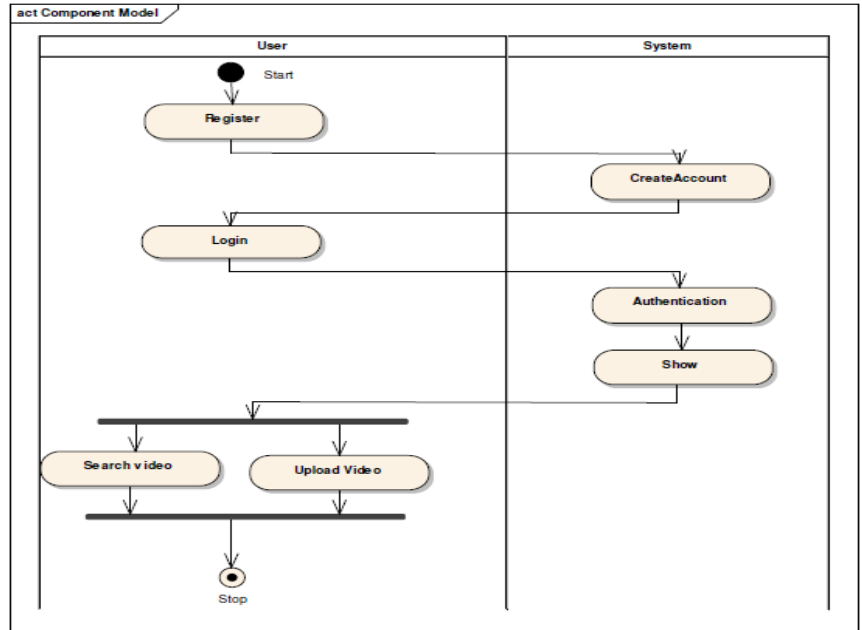

Figure 5. Activity diagram for User

### 3.4 StateChart Diagram

A Statechart diagram describes the states machine. Now to clarify it state machine can be defined as a machine which defines different states of an object and these, is a special kind of a State transition diagram. As Statechart diagram defines states it is used to model lifetime of an object. Figure 6. Shows the states for Admin and Figure 7. Shows the state of User.

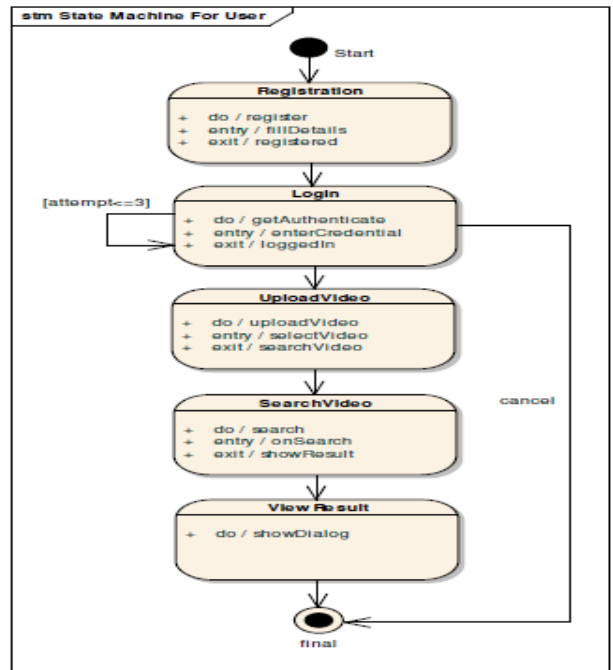

Figure 6. Statechart diagram for User

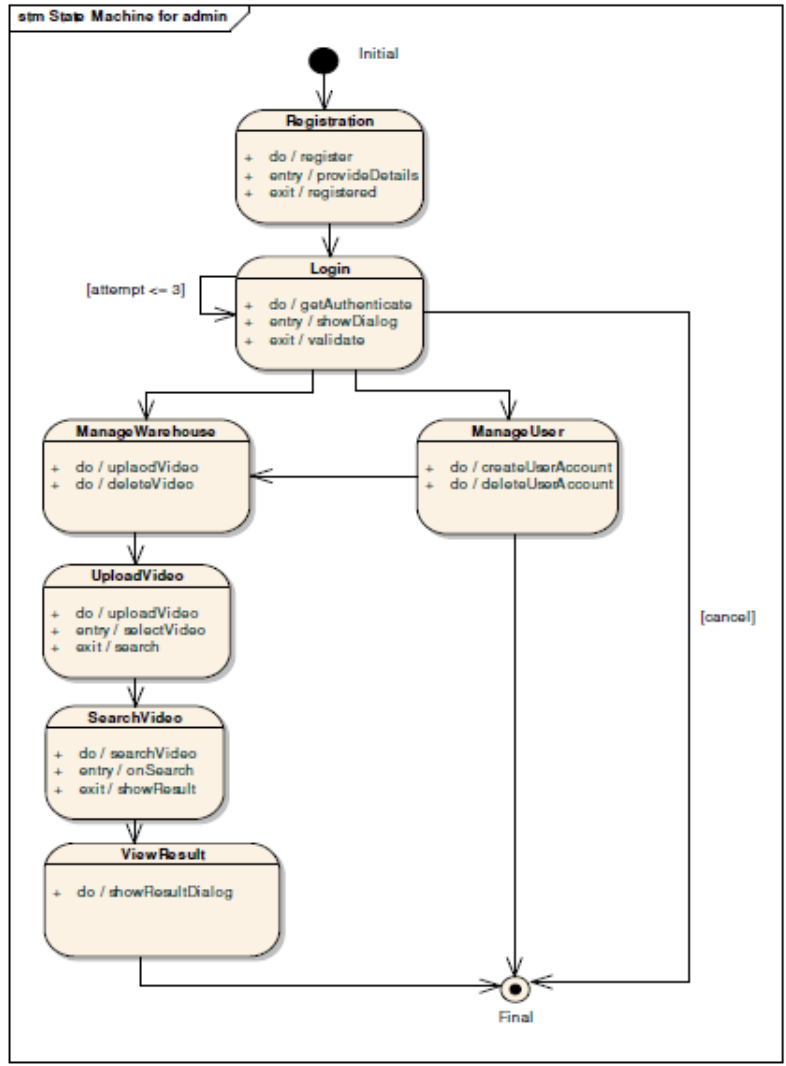

Figure 7. Statechart diagram for Admin

# 4. SYSTEM DESIGN

## 4.1 Component Diagram

It models the physical components in design the way of looking at the components is the packages . A package is used to show how we can group together classes , which in essence are smaller scale components .

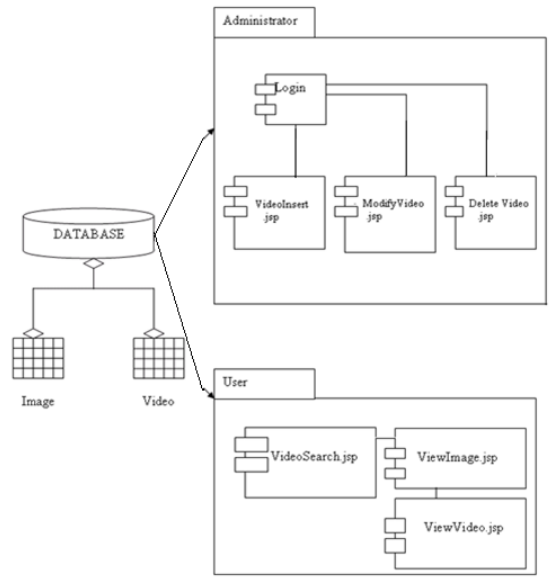

Figure 8. Component diagram

Here administrator manages login, addition, modification and deletion of video from database and user searches video based on index frame and views similar videos as shown in figure 8.

## 4.2 Deployment Diagram

Deployment Diagram shows the configuration of runtime processing elements and the software components, processes and objects that live in them. It is shown in figure 6. A software component instance represents runtime manifestations of code units. In most cases, component diagrams are used in the conjunction with deployment diagram to show how physical modules of code are distributed on various hardware platforms. In many cases, component and deployment diagrams can be combined. A deployment diagram is graph of nodes connected by communication association. Nodes may contain a component instance, which means that component lives or runs at that mode. Components may contain objects; this indicates that the object is a part of component. Components are connected to other components by dashed arrow dependencies, usually through interfaces, which indicate one component uses service of another. Each node or processing element in the system represented by a three dimensional box. Connection between the node (or platforms) themselves are shown by solid lines.

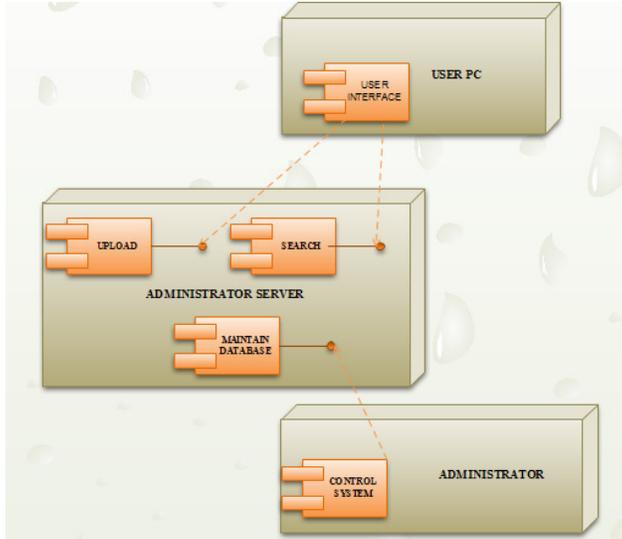

Figure 9. Deployment Diagram

## 4.3 Database Design

Database design is used to define and then specifies the structure of business objects used in the client/server system. The analysis required to identify business objects is accomplished using business process engineering methods. Conventional analysis modeling notation such as the ERD, can be used to define business objects, but a database repository should be established in order to capture the additional information that cannot be fully documented using a graphic notation such as an ERD.

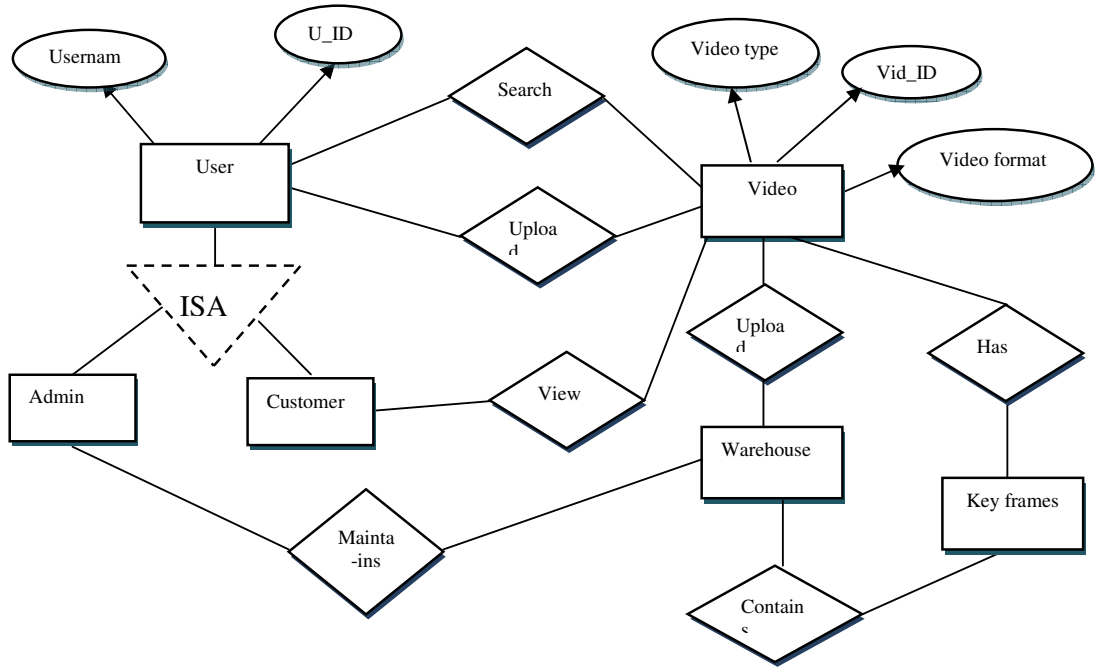

Figure 10. E-R Diagram for Video Search Engine

In this repository, a business object is defined as information that is visible to the purchasers and users of the system, not its developers. This information, implemented using a relational database, can be maintained in a design repository. Following are the database tables used for implementing proposed systems on which queries can be performed.

Video_store(v_id, v_name, video,stream)
Key_frames(v_id,i_name,image,min,max,sch,gabor,glcm,tamura,majorRegions,v_id)
Where v_id - video id,
    v_name - name of video file,
    video - video file,
    stream - stream of keyframes,
    i_id – video frame id,
    min-max - indexing range,
    sch – simple colour histogram,
    gabor, glcm, tamura - feature texture string,
    majorRegion - no. of max regions

Using following Oracle query   Video_store table is created for indexing and retrieval of video.

```
CREATE TABLE  "VIDEO_STORE"
  ( "V_ID" NUMBER NOT NULL ENABLE,
    "V_NAME" VARCHAR2(60),
    "VIDEO" ORD_ Video,
    "STREAM" BLOB,
    "DOSTORE" DATE,
    PRIMARY KEY ("V_ID") ENABLE   )
```

Using following Oracle query  KEY_FRAMES table is created for storing key frames and their features.

```
CREATE TABLE  "KEY_FRAMES"
  ( "I_ID" NUMBER NOT NULL ENABLE,
    "I_NAME" VARCHAR2(40) NOT NULL ENABLE,
    "IMAGE" ORD_ Image,
    "MIN" NUMBER,
    "MAX" NUMBER,
    "SCH" VARCHAR2(1500),
    "EDGEDENSITY" VARCHAR2(250),
    "MAJORREGIONS" NUMBER,
    "V_ID" NUMBER,
    PRIMARY KEY ("I_ID") ENABLE    )
```

## 4.4 Algorithm Design

In this section we present the various algorithms used for the implementing proposed content based video search engine.

### 4.4.1 Key frame extraction Algorithm

It works on group of frames extracted from a video. It takes a list of files or frames in order in which they will be extracted. It is based on predefined threshold that specify whether 2 video frame are similar. It s main function is to choose smaller number of video representative key frames.

   i.   Starts from 1st frame from sorted list of files.
   ii.   If consecutive frames are within threshold, then two frames are similar.
   iii.   Repeat process till frames are similar, delete all similar frames & take 1st as key -frame.
   iv.   Start with next frame which is outside threshold & repeat the steps i to a for all frames of video .

**Input -** Frames of video extracted by video to jpeg converter.
**Output -** Key frames with less similarity & representing video.

**Pseudo code for key frame extraction**
1] Let i & j be integer variables, ri1 & ri2 be RenderedImage objects.
2] Get all Jpeg files in files array.
3] Initialize length to length of files array.
4] Do
a] i=0
    b] while i is not equal to length
          i] Create RenderedImage ri1 object that hold rescaled IVersion of image file i.
          ii] Do
                a] j=i+1
                b] while j is not equal to length
                c] Create RenderedImage ri2 object that holds rescaled IVersion of image file j.
                d] dist = difference between ri1 & ri2.
                e] if(dist > 800.0)
                      { i=j-1; break; }
                  else
                    delete file j.
                f] j=j+1
    c] i=i+1
5] End

**4.4.2 Histogram based range finder indexing algorithm**

We implemented tree based indexing algorithm consisting of  grouping frames into pixel ranges such as 0-255 on first level, 0-127 or 128-255 on second level so on third, fourth level. Algorithm level by level calculates min – max range finally if when thresholding criteria is not satisfied, frame is grouped onto previous level into min –max range. Every frame is part of first level as it satisfies thresholding criteria.

1. Calculate histogram of image.
2. Calculate min–max range by grouping pixel count.
3. Find if thresholding criteria are satisfied.
4. Go to next lower level & repeat step from 2 to 4.
5. If criteria are not satisfied, image is grouped on previous level where min – max range is satisfied as shown in figure 11.

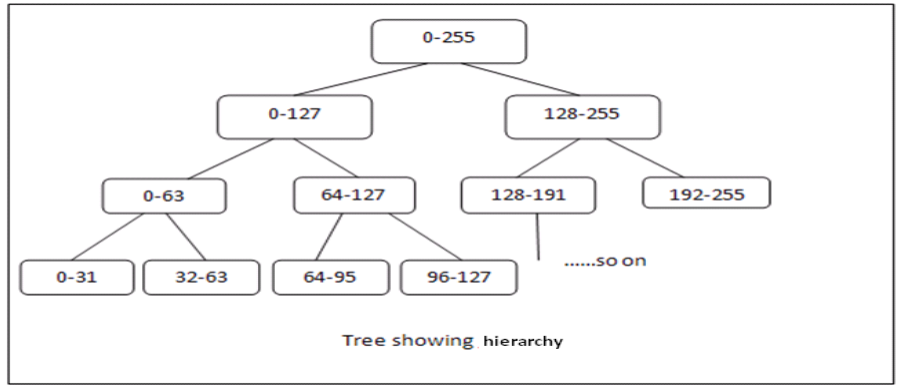

Figure 11. Indexing tree

**Input :** Video frame after choosing key frames as image.
**Output :** min – max range that groups image in particular group.

**Pseudo code for indexing process :**
Variables min & max store range for indexing process.
1] Let min = 0,max = 255;
2] Let long sum = 0;
3] Let double result = 0;
4] //1st block test
    A] i=min
    B] while i is not equal to max
                i] sum = sum + histogram[i]
    C] result = sum/900.0;
        D] if(result>55.0)
                {min = 0;max = 127;}
           else
        {min = 128;max = 255;}
    E] i=i+1

5] //2nd block test
    sum = 0, result = 0;
6] if(min = = 0 && max = = 127)
      {
        A] i=0
    B] while i is not equal to 63
                i] sum = sum + histogram[i]
    C] result = sum/900.0;
    D] if(result>60.0)
        {min = 0;max = 63;}
         else
        {
              I] sum = 0, result = 0;
              II] for (int i = 64; i < 127; i++)
                    a] sum = sum + histogram[i];
              III] result = sum/900.0;
              IV] if(result>60.0)
                  {min = 64;max = 127;}
        }
      }
7] sum = 0, result = 0;

```
8] if(min = = 128 && max = = 255)
             {
    A] for (int i  =  128; i < 191; i++)
                      i] sum = sum + histogram[i];
         B] result = sum/900.0;
         C] if(result>60.0)
                      {min = 128;max = 191;}
            else
            {
                      I] sum = 0, result = 0;
                      II] for (int i  =  192; i < 255; i++)
                                a] sum = sum + histogram[i];
                      III] result = sum/900.0;
                      IV] if(result>60.0)
                                   {min = 192;max = 255;}
            }
         }

9] //3rd block test
     sum = 0, result = 0;
10] if(min = = 0 && max = = 63)
             {
    A] for (int i  =  0; i < 31; i++)
                      i] sum = sum + histogram[i];
         B] result = sum/900.0;
         C] if(result>60.0)
                      {min = 0;max = 31;}
            else
            {
                      I] sum = 0, result = 0;
         II] for (int i  =  32; i < 63; i++)
                      a] sum = sum + histogram[i];
         III] result = sum/900.0;
                   IV] if(result>60.0)
                                   {min = 32;max = 63;}
            }
         }

11] sum = 0, result = 0;
12] if(min = = 64 && max = = 127)
                {
    A] for (int i  =  64; i < 95; i++)
                      i]sum = sum + histogram[i];
      B] result = sum/900.0;
      C] if(result>60.0)
                      {min = 64;max = 95;}
                else
                   {
         I] sum = 0, result = 0;
         II] for (int i  =  96; i < 127; i++)
                             a]sum = sum + histogram[i];
                      III] result = sum/900.0;
                      IV] if(result>60.0)
                                   {min = 96;max = 127;}
                   }
                }

13] sum = 0, result = 0;
```

14] if(min = = 128 && max = = 191)
        {
  A] for (int i  =  128; i < 159; i++)
           i]sum = sum + histogram[i];
  B] result = sum/900.0;
  C] if(result>60.0)
           {min = 128;max = 159;}
      else
        {
      I] sum = 0, result = 0;
      II] for (int i  =  160; i < 191; i++)
          a] sum = sum + histogram[i];
      III] result = sum/900.0;
      IV] if(result>60.0)
            {min = 160;max = 191;}
        }
    }

15] sum = 0, result = 0;
16] if(min = = 192 && max = = 255)
      {
  A] for (int i  =  192; i < 223; i++)
         a]sum = sum + histogram[i];
  B] result = sum/900.0;
  C] if(result>60.0)
          {min = 192;max = 223;}
      else
        {
      I] sum = 0, result = 0;
      II] for (int i  =  224; i < 255; i++)
          a] sum = sum + histogram[i];
      III] result = sum/900.0;
      IV] if(result>60.0)
        {min = 224;max = 255;}
    }
    }

### 4.4.3 GLCM texture Feature Extraction

The Gray Level Co-occurrence Matrix1 (GLCM) and associated texture feature calculations are image analysis techniques. Given an image composed of pixels each with an intensity (a specific gray level), the GLCM is a tabulation of how often different combinations of gray levels co-occur in an image or image section.

**Input :** image as PlanarImage object
**Output** : string containing feature values.

**Pseudo code for GLCM texture:**
public GLCM_Texture(PlanarImage im, boolean preprocess):
1] If(preprocessor is true) input = preprocess the image in im;
2] otherwise input = im;
3] create a data structure Raster for input image
4] initialize k=0 ,h=o ,w=0;
    i] while h is not equal to height of image
        a] while i is not equal to width of image ;
        A] Initialise pixels[w][h] with value for pixel in image.
        B] Initialise pixel[k++] with value for pixel in image.

**private** PlanarImage preprocess(PlanarImage input)
{
    **1] if** (input has IndexColorModel)
        a] Retrieve the IndexColorModel of image;
        b] Cache the number of elements in each band of the colormap in mapSize integer
          variable.
        c] Allocate an array for the lookup table data i.e. lutData = **new byte**[3][mapSize];
        d] Load the lookup table data from the IndexColorModel.
          i] Get Red in lutData[0] from IndexColorModel object ;
          ii] Get Green in lutData[1] from IndexColorModel object ;
          iii] Get Blue in lutData[2] from IndexColorModel object ;
        e] Create the lookup table object i.e. LookupTableJAI  using lutData array;
        f] Replace the original image i.e. input with the 3-band RGB image using lut object.

    2] **if** ( NumBands in input > 1)
        a] Let matrix = {{ 0.114, 0.587, 0.299, 0 }};
        b] Replace the original image i.e. input with bandcombineb using matrix;

    **3]return** input;

public void calcTexture()
        1] Let a = 0 be integer variables;
        2] Let b = 0, y=0, x=0 be integer variables;
        3] Let offset and i be integer variables;
        4] Let glcm be a 2-D array of 257 by 257;
        5] while y is not equal to height
          i]offset = y*width;
        6] while x is not equal to width
          i] i = offset + x;
          ii] a = 0xff & pixel[i];
          iii] b = 0xff & pixels[x+*step*][y];
          iv] glcm [a][b] +=1;
          v] glcm [b][a] +=1;
          vi] pixelCounter +=2;
        7] while a is not equal to 257
          i] while b is not equal to 257
              a] glcm[a][b]= glcm[a][b] / pixelCounter;
        8] while a is not equal to 257
          i] while b is not equal to 257
              a] asm= asm+ (glcm[a][b]*glcm[a][b]);
        9] Initialize double px= 0;
        10] Initialize double py= 0;
        11] Initialize double meanx= 0.0;
        12] Initialize double meany= 0.0;
        13] Initialize double stdevx= 0.0;
        14] Initialize double stdevy= 0.0;
        15] while a is not equal to 257
          i] while b is not equal to 257
              a] px= px+a*glcm [a][b];
              b] py= py+b*glcm [a][b];
        16] while a is not equal to 257
          i] while b is not equal to 257
              a] stdevx= stdevx+(a-px)*(a-px)*glcm [a][b];
              b] stdevy= stdevy+(b-py)*(b-py)*glcm [a][b];
        17] while a is not equal to 257
          i] while b is not equal to 257
              a] correlation= correlation+ (a-px)*(b-py)*glcm [a][b]/(stdevx*stdevy);

     18] while a is not equal to 257
       i] while b is not equal to 257
          a] IDM=IDM+ glcm[a][b]/(1+(a-b)*(a-b))
     19] while a is not equal to 257
       i] while b is not equal to 257
          a] if (glcm[a][b]==0) { }
              b] else
              entropy = entropy - glcm[a][b] * $log$(glcm[a][b]) ;

public String getStringRepresentation()
    1] Build new object StringBuilder sb;
    2] Append the string GLCM texture to sb;
    3] Append pixelCounter to sb;
    4] Append asm to sb;
    5] Append contrast to sb;
    6] Append correlation to sb;
    7] Append IDM to sb;
    8] Append entropy to sb;
    9] Return string format of sb;

## 4.4.4 Color histogram Extraction

The color space of frame is quantizes into a finite number of discrete levels. Each of this level becomes bin in the histogram. The color histogram is then computed by counting the number of theses discrete levels. Color Histogram(h) is then computed with the color information like $h_r(i)$, $h_g(i)$, $h_b(i)$ to represent the color domain.

**Input :** image as BufferedImage object
**Output** : string containing histogram values.
**pseudo code for color histogram :**
public void extract(BufferedImage image)
    1] Let pixel be 2-D array of size width & height.
    2] Create a data structure Raster for input image.
    3] x = 0  , y=0 & while x is not equal to width of image
       A] while y is not equal to height of image
      i] pixel[x][y]= corresponding value in raster.
         ii] if(histogramType != HistogramType.*HMMD*) histogram[quant(pixel)]++.

 public String getStringRepresentation()
    1]  Create a StringBuilder object.
    2] Append histogramType to sb.
    3] Append  ' ' to sb.
    4] Append length of  histogram to sb.
    5] Append  ' ' to sb.
    6] for (int i = 0; i < length of  histogram. i++)
       A] Append histogram[i] to sb.
       B] Append  ' ' to sb.
    7] Return string format of sb.

## 4.4.5 Motion Feature Extraction

**Input :** Sketch with motion trajectory
**Output :** videos consisting of the same motion stroke given in input.

**Pseudo-code :**
1.  Keyframe selection

2. Keyframe pre-processing
3. Acquire the drawn trajectory and save its co-ordinate vectors.
4. To find min. Distance for object
5. Body sectioning along with the trajectory
6. Determine the object index for the trajectory.
7. Trajectory mapping onto body
8. Gradient calculations :
Calculate the directional gradient of the trajectory along the co-ordinate axes.
     i) The rate of change in the spatial dimension is calculated along the axes.
     ii) Above value is converted into discrete form for the comparison purpose.
9. Gradient mapping and correlation(binarisation and Rate of change of gradient) i.e. The co-relation between the indexed gradient and the current trajectory is calculated.
10. Based on the match between the above two components, rank is assigned to the videos.
11. Video retrieval using indexes.

# 5. EXPERIMENTAL RESULT

The proposed retrieval system is implemented using Java programming language and tested using the database video clips of MPEG-2 format. Database contains collection of 100 videos out of which 40 video clips are of sports category, 15 are of movie category, 15 is of e-learning catagory and 30 are animated movies. The retrieval results are shown in table 1. Sample videos were downloaded from www.archive.org, www.gettyimages.com and www.youtube.com. All results are taken by using Personal computer Intel i3-2120 processor with 4 GB Memory. Time analysis results may vary by different computers.

| Query Video | Most Matched | Total videos Retrieved by our system | Similar Available in Database |
|---|---|---|---|
| vid1.mpeg | 4 | 5 | 9 |
| vid2.mpeg | 5 | 5 | 7 |
| vid3.mpeg | 4 | 5 | 8 |
| vid4.mpeg | 8 | 9 | 12 |
| vid5.mpeg | 8 | 11 | 14 |
| vid1.mpeg | 9 | 11 | 12 |
| vid2.mpeg | 8 | 8 | 10 |
| vid3.mpeg | 7 | 8 | 9 |
| vid4.mpeg | 7 | 7 | 10 |
| vid5.mpeg | 8 | 8 | 9 |

Table1. Results for most matched videos and retrieve videos from database

| Query Video | Time needed for retrieval (in Sec) | Time needed for similarity matching (in Sec) |
|---|---|---|
| vid1.mpeg | 0.15 | 0.02 |
| vid2.mpeg | 0.48 | 0.32 |
| vid3.mpeg | 0.31 | 0.28 |
| vid4.mpeg | 0.89 | 0.02 |
| vid5.mpeg | 0.18 | 0.25 |
| vid1.mpeg | 0.17 | 0.03 |
| vid2.mpeg | 0.58 | 0.42 |
| vid3.mpeg | 0.41 | 0.24 |
| vid4.mpeg | 0.95 | 0.04 |
| vid5.mpeg | 0.28 | 0.15 |

Table2. Time analysis for similarity matching by ED and video retrieval from database.

$$Precision = \frac{number\ of\ videos\ retried\ relevant\ to\ query\ video}{Total\ number\ of\ videos}$$

$$Recall = \frac{number\ of\ videos\ retried\ relevant\ to\ query\ video}{Total\ number\ of\ available\ videos\ relevant\ to\ query\ clip}$$

| Query Video | Precision | Recall |
|---|---|---|
| vid1.mpeg | 0.80 | 0.56 |
| vid2.mpeg | 1.00 | 0.71 |
| vid3.mpeg | 0.80 | 0.63 |
| vid4.mpeg | 0.89 | 0.75 |
| vid5.mpeg | 0.73 | 0.79 |
| vid1.mpeg | 0.82 | 0.92 |
| vid2.mpeg | 1.00 | 0.80 |
| vid3.mpeg | 0.88 | 0.89 |
| vid4.mpeg | 1.00 | 0.70 |
| vid5.mpeg | 1.00 | 0.89 |

Table3: precision and recall for each query video

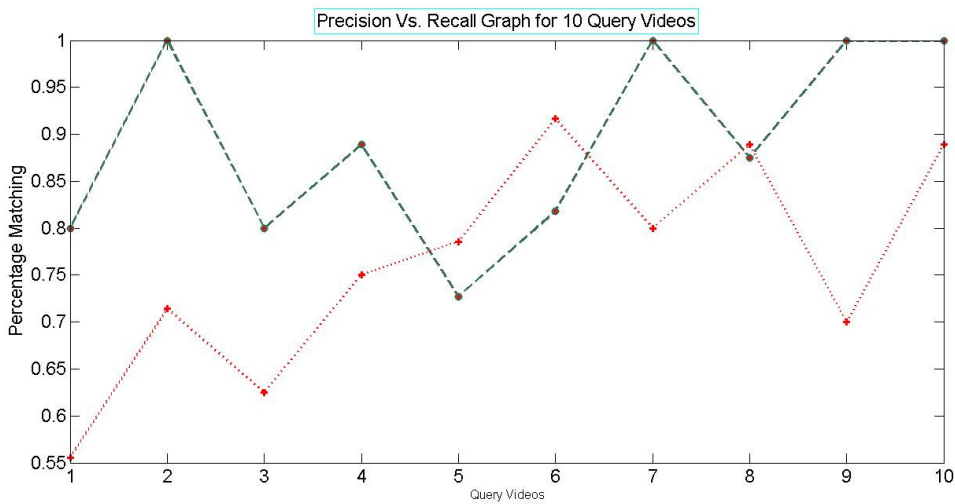

Fig 12. Precision recall graph for query videos

# 6. CONCLUSION

Video Retrieval is broad area that integrates features from several features and fields including artificial intelligence, machine learning, data base management systems, etc. There have been large numbers of algorithms rooted in these fields to perform various video retrieval tasks. In our proposed system we have implemented retrieval system by integrating various features of query video frame. Experimental results show that integration of extracted features improves video indexing and retrieval. This is demonstrated by the finding that multiple features produces effective and efficient system as precision and recall values are improved.

## REFERENCES

[1]  G Eason, Weining Hu, ,NianhuaXie,LI LI,Xianglin Zeng,Stephan Maybank"A survey on visual content-based video indexing and retrieval",IEEE Transactions on systems,man,and cybernetics-part c:Applications and Reviews,Vol 41,No.6,November 2011.

[2]  Choi, K.-C. Ko, Y.- M. Cheon, G-Y. Kim, H-Il, S.-Y. Shin, and Y.-W.Rhee, "Video shot boundary detection algorithm," Comput. Vis.,Graph. Image Process., (Lect. Notes Comput. Sci.), 4338, pp. 388–396,2006.

[3]  K.Mandal,F Idris,S Panchanathan, "Image and Video indexing in the Compressed Domain", to apear in Image and Vision Computing Journal, DOI 10.1.1.40.8228

[4]  Cao Zheng, Zhu Ming, "An Efficient  Video Similarity Search Strategy For Video-On-Demand Systems"DOI 978-1-4244-4591-2 IEEE,Proceedings of IC-BNMT 2009

[5]  Kebin Jia,Xiuxin Chen, "A Video Retrieval Algorithm Based on Spatio-temporal Feature Curves" IEEE International Conference on MultiMedia and Information Technology, DOI 978-0-7695-3556-2/08 $25.00 © 2008

[6]  H. J. Zhang, J. Wu, D. Zhong, and S. W. Smoliar, "An integrated system for content-based video retrieval and browsing," Pattern Recognit.,vol. 30, no. 4, pp. 643–658, 1997.

[7]  Daga Brijmohan, Avinash Bhute, and Ashok Ghatol. "Implementation of Parallel Image Processing Using NVIDIA GPU Framework." Advances in Computing, Communication and Control. Springer Berlin Heidelberg, 2011. 457-464.

[8]  Bhute Avinash N., and B. B. Meshram. "Automated Multimedia Information Retrieval Using Color and Texture Feature Techniques." IJECCE 3.5 (2012): 885-891.

[9]  Bhute Avinash N., B. B. Meshram, and Harsha A. Bhute. "Multimedia Indexing and Retrieval Techniques: A Review." International Journal of Computer Applications 58.3 (2012): 35-42.

[10] G. Quenot, D. Moraru, and L. Besacier. (2003). "CLIPS at TRECVID: Shot boundary detection and feature detection,"in Proc. TREC Video Retrieval Eval. Workshop Notebook Papers [Online].http://www-nlpir.nist.gov/projects/tvpubs/tv.pubs.org.html#2003

[11] Liu David and Chen Tsuhan, "Video retrieval based on object discovery", Computer vision and image understanding, Vol.113, No.3, pp.397-404, 2009.

[12] Petkovic, Milan, Jonker, Willem,"Content-based video retrieval", Kluwer Academic Publishers, Boston, Monograph, 2003, 168 p., Hardcover ISBN: 978-1-4020-7617-6

[13] Che-Yen Wen, Liang-Fan Chang and Hung-Hsin Li, "Content based video retrieval with motion vectors and the RGB color model", Forensic Science Journal, Vol.6, No.2, pp .1-36,  2007.

[14] W. M. Hu, D. Xie, Z. Y. Fu, W. R. Zeng, and S. Maybank, "Semanticbased surveillance video retrieval," IEEE Trans. Image Process., vol. 16,no. 4, pp. 1168–1181, Apr. 2007.11

[15] P. Browne and A. F. Smeaton, "Video retrieval using dialogue, keyframe similarity and video objects," in Proc. IEEE Int. Conf. Image Process.,Sep. 2005, vol. 3, pp. 1208–1211.

[16] Y. Wu, Y. T. Zhuang, and Y. H. Pan, "Content-based video similarity model," in Proc. ACM Int. Conf. Multimedia, 2000, pp. 465–467.3

[17] C. G. M. Snoek, B. Huurnink, L. Hollink, M. de Rijke, G. Schreiber, and M. Worring, "Adding semantics to detectors for video retrieval," IEEE Trans. Multimedia, vol. 9, no. 5, pp. 975–985, Aug. 2007.

[18] S. Y. Neo, J. Zhao,M. Y. Kan, and T. S. Chua, "Video retrieval using high level features: Exploiting query matching and confidence-based weighting," in Proc. Conf. Image Video Retrieval, Singapore, 2006, pp. 370–379.

[19] Y. Aytar, M. Shah, and J. B. Luo, "Utilizing semantic word similarity measures for video retrieval," in Proc. IEEE Conf. Comput. Vis. Pattern Recog., Jun. 2008, pp. 1–8.16

[20] R. Yan and A. G. Hauptmann, "A review of text and image retrieval approaches for broadcast news video," Inform. Retrieval, vol. 10, pp. 445–484, 2007

[21] Liang-Hua Chen,Kuo-hao Chin,Hong-Yuan Liao,"An Integrated Approach to Video Retrieval",Nineteenth Australasian Database Conference(ADC2008), Wollongong, Australia,Vol 75 January 2008

[22] Chung Wing Ng, Irwin King and Michael R. Lyu, "Video Comparison Using Tree Matching Algorithms"in Proc Conf Audio,Speech and Launguage processing,vol 19,No. 1,pp. 196-205,Jan 2011

[23] Young-tea,Kim,Tat-Sang,Chua, "Retrieval of News Video using video sequence" , Proceedings of 11th International Multimedia Modelling Conference (MMM'05),1550-5502/05

[24] R Datta, D Joshi, J Li, and J. Wang, (2008) "Image Retrieval: Ideas, Influences, and Trends of the New Age", ACM Computing Surveys, VOl 40, No. 2, April 2008.

[25] Tristan Glatard, Johan Montagnat, Isabelle E. Magnin (2004)," Texture Based Medical Image Indexing and Retrieval: Application to Cardiac Imaging", MIR'04, October 15–16, 2004, New York, New York.

[26] Egon L. van den Broek, Peter M. F. Kisters, and Louis G. Vuurpijl (2004)," Design Guidelines for a Content-Based Image Retrieval Color-Selection Interface"ACM Dutch Directions in HCI, Amsterdam.

[27] Tristan Glatard, Johan Montagnat, Isabelle E. Magnin (2004), "Texture Based Medical Image Indexing and Retrieval: Application to Cardiac Imaging", ACM Proc. Of MIR'04, October 15–16, 2004, New York, New York, USA.

[28] H. Baird, D. Lopresti, B. Davison, W. Pottenger (2004), "Robust document image understanding technologies", HDP, ACM.

[29] R. Datta D. Joshi, J. LI., JAMES Z. W., "Image Retrieval: Ideas, Influences, and Trends of the New Age", ACM Transactions on Computing Surveys, Vol. , No. , 20, pp. 1-65.

[30] Jing-Fung Chen,, Hong-Yuan Mark Liao1, and Chia-Wen Lin (2005)," Fast Video Retrieval via the Statistics of Motion Within the Regions-of-Interest", KES'05 Proceedings of the 9th international conference on Knowledge-Based Intelligent Information and Engineering Systems - Volume Part III Springer.

[31] Xi Li, Weiming Hu, Zhongfei Zhang, Xiaoqin Zhang, Guan Luo (2008), "Trajectory-Based Video Retrieval Using Dirichlet Process Mixture Models", 19th British Machine Vision Conference (BMVC).

[32] Lin Lin; Chao Chen; Mei-Ling Shyu; Shu-ChingChen "Weighted Subspace Filtering and Ranking Algorithms for Video Concept Retrieval", IEEE Multimedia, Volume: 18 , Issue: 3, 2011.

[33] Chih-Chin Lai; Ying-Chuan Chen, "A User-Oriented Image Retrieval System Based on Interactive Genetic Algorithm", IEEE Transactions on Instrumentation and Measurement, Volume: 60 , Issue: 10, 2011.

[34] Shigeaki Sakurai, Hideki Tsutsui, "A clustering method of bloggers based on social annotations", International Journal of Business Intelligence and Data Mining, Volume 6 Issue 1, 2011.

[35] Masami Shishibori, Daichi Koizumi, Kenji Kita, "Fast retrieval algorithm for earth mover's distance using EMD lower bounds and a skipping algorithm", Advances in Multimedia, Volume 2011

[36] Ballan, L.; Bertini, M.; Del Bimbo, A.; Serra, G, " Video Annotation and Retrieval Using Ontologies and Rule Learning", IEEE Multimedia, Volume: 17 , Issue: 4,: 2010.

[37] Beecks, C.; Uysal, M.S.; Seidl, T., "A comparative study of similarity measures for content-based multimedia retrieval", IEEE International Conference onMultimedia and Expo (ICME), 2010